\documentclass[aps,prb,twocolumn,groupedaddress]{revtex4-1}
\usepackage{graphicx}
\usepackage{mathtools}
\usepackage{amssymb}
\usepackage{color}
\usepackage{float}

\DeclareFontFamily{U}{mathb}{\hyphenchar\font45}
\DeclareFontShape{U}{mathb}{m}{n}{
      <5> <6> <7> <8> <9> <10> gen * mathb
      <10.95> mathb10 <12> <14.4> <17.28> <20.74> <24.88> mathb12
}{}
\DeclareSymbolFont{mathb}{U}{mathb}{m}{n}
\DeclareMathSymbol{\rightsquigarrow}{3}{mathb}{"F9}

\begin{document}

\title{Evaporating black holes have leaky horizons or exotic atmospheres}

\maketitle

\noindent
Samuel L.\ Braunstein and Stefano Pirandola\\
Computer Science, University of York, York\\
${~}$YO10 5GH, UK\\

\vskip -0.1truein
\noindent
{\bf 
Classically, black holes are compact objects with perfect semi-permeable
horizons: Anything may enter, nothing may leave.  We consider an
axiomatic approach that applies to any black hole type, including
arbitrarily near-extremal black holes, that can unitarily evaporate away
completely by any mechanism. We show that a quantum black hole must
either have a leaky horizon (allowing quantum information out), or it
must look very different from its classical counterpart having an
external neighborhood consisting of exotic matter with super-ordinary
entropic content (such as an `atmosphere' of microscopic black holes).
}
\vskip 0.1truein

The black hole information paradox \cite{Hawking76} points to a
fundamental clash between gravity and quantum mechanics. The paradox
relies centrally on the mechanism by which a black hole evaporates and
was derived specifically for pair creation into vacuum \cite{Hawking76}.
Pair creation has remained the most fundamental and hence most well
accepted mechanism for black hole evaporation.  However, the discovery
of the black hole `energetic curtain' \cite{Braunstein13} (also known as
a `firewall' \cite{Mathur09,AMPS}) from the Page time onwards (when
one-half the black hole's initial entropy has evaporated away) casts
doubt on the validity of pair creation into vacuum as the universal
mechanism by which a black hole evaporates for the vast majority of its
lifetime (see Ref.~\onlinecite{Braunstein07} for earlier evidence and the
Supplementary Information for an alternative argument).

Here we recast the information paradox in an axiomatic manner that does
not rely on any specific evaporation mechanism. Our approach allows us
to do away with many conventional assumptions, e.g., the need to i)
quantify information content beyond measures of entropy; ii) rely at all
on the quantum state within the black hole; iii) rely on a finite
dimensionality for the black hole interior Hilbert space; or iv) rely on
physics at the Planck scale.  We begin by stating three, widely accepted
axioms:

{\bf I.\ Unitary evolution}. For unitarity to be preserved during black
hole evaporation, a black hole cannot stop evaporating when  it reaches
some `stable remnant' \cite{Hooft,Giddings95,Bekenstein94}, nor can its
interior `bud off' as a baby universe \cite{Banks84,Hooft}. Thus, this
axiom implies that black holes evaporate away completely and unitarily.

{\bf II.\ Black holes are classically causal}. The causal structure of a
classical black hole implies that the event horizon of a quantum black
hole is a perfect semi-permeable membrane \cite{Hawking76}. Our argument
will not rely on this property for Planck-scale black holes.

{\bf III.\ Weak correspondence principle}. To reconcile gravity with
quantum mechanics, the environment of a black hole must look alike,
whether classical or quantum mechanical.  To enforce an exact
correspondence, black holes are typically assumed to evaporate into
vacuum (as seen by an infalling observer). Our third axiom makes a
weaker statement that externally, quantum black holes do not look too
different from their theoretical classical counterparts. In particular,
the region surrounding a quantum black hole does not consist of exotic
matter with super-ordinary entropic content (such as an `atmosphere' of
microscopic black holes).

We now show that the above three axioms are inconsistent: The presence
of an ideal horizon (II) leads either to the failure of complete unitary
evaporation (violating I) or to an exotic atmosphere (violating III).

{\it Classically-causal radiation production:}---Although our analysis
applies to any unitary radiation production mechanism we shall
specifically describe it in the context of black hole radiation. First,
we associate our process with some specific black hole and presume that
to an excellent approximation radiation is not produced arbitrarily far
away. In other words, we suppose that there is some finite region
encompassing the black hole within which the radiation is generated.

\begin{figure}[ht]
\includegraphics[width=75mm]{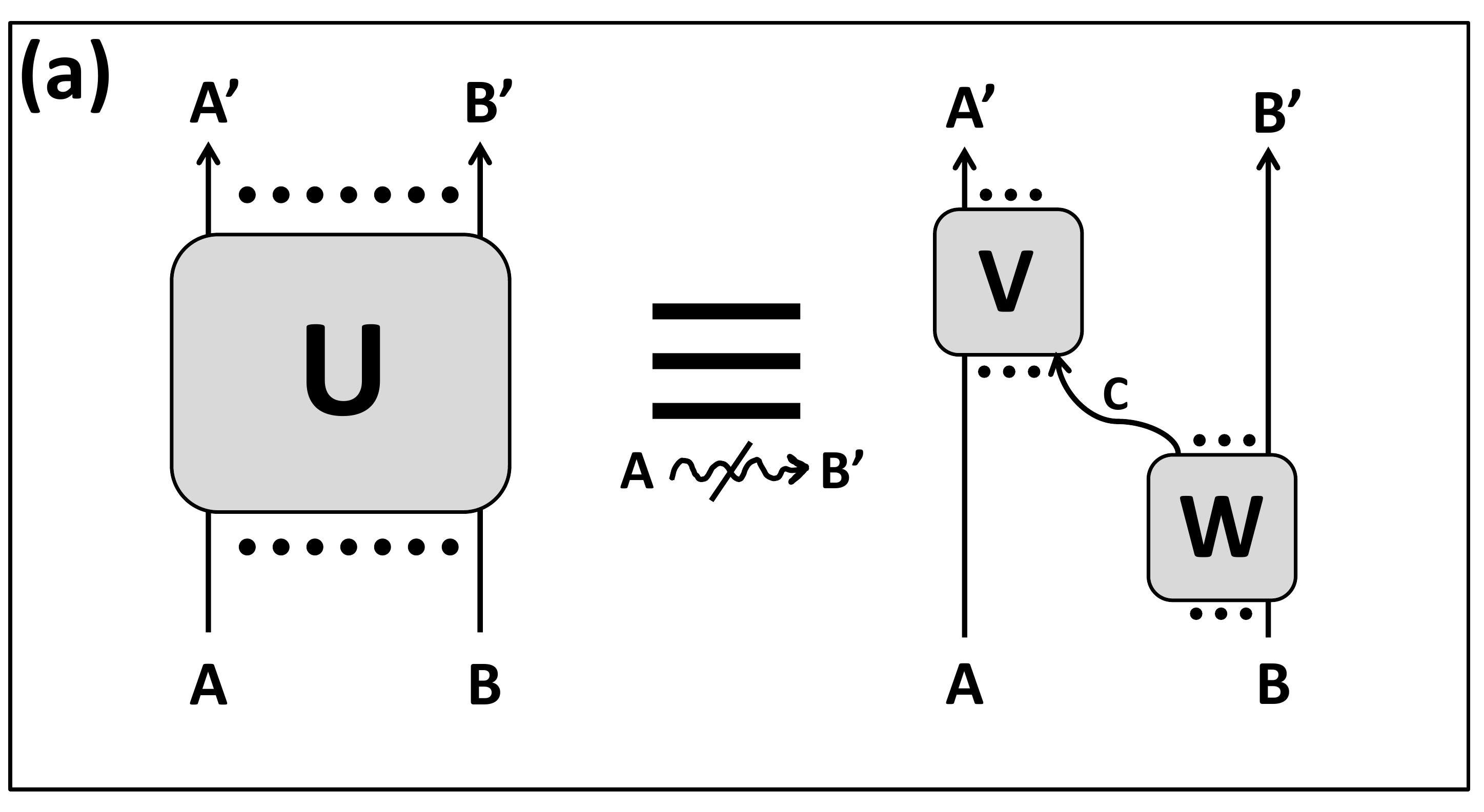}
\vskip -0.15truein
\includegraphics[width=75mm]{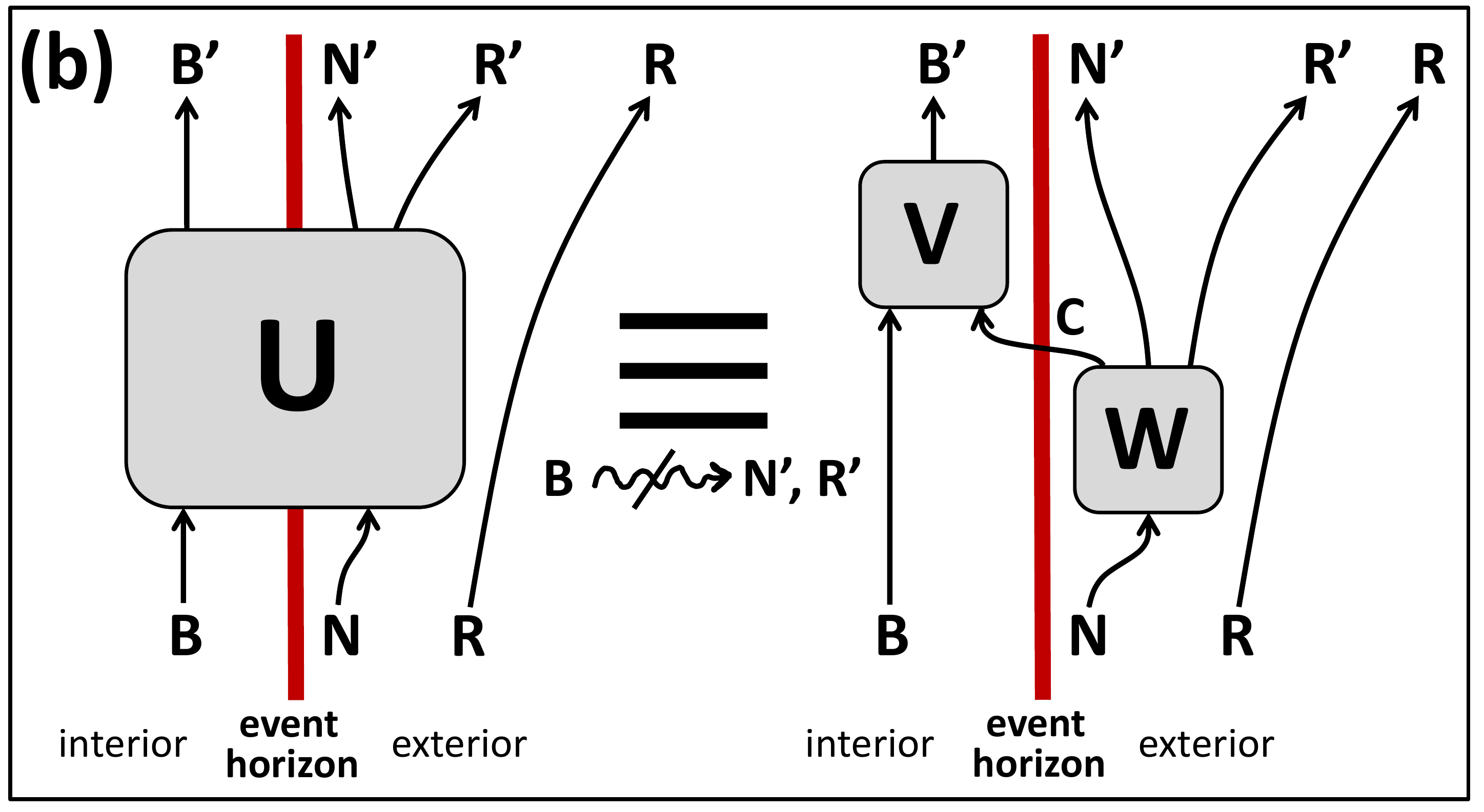}
  \caption{
(a) Quantum circuit diagram of the no-communication decomposition
theorem \cite{Werner}: Any unitary process $U$ (left-hand circuit) which
maps subsystems $A$ and $B$ into $A'$ and $B'$ but which does not allow
communication from $A$ to $B'$, $A\mathrlap{\kern
0.6em\not}\rightsquigarrow B'$, can be decomposed into a pair of unitary
subprocesses $V$ and $W$ (right-hand circuit) possibly connected by a
`reverse communication' channel $C$.  (The dots refer to any ancillary
degrees-of-freedom.)
(b) Schematic generation of radiation $R'$ during some epoch. $N$ and
$N'$ are the degrees-of-freedom in the exterior neighborhood prior and
posterior to the unitary operation, respectively; $B$ and $B'$ label
interior degrees-of-freedom; and $R$ denotes radiation from an earlier
epoch. Assuming no communication across the event horizon from interior
to exterior, we may decompose the unitary process $U$ into two
subprocesses $V$ and $W$ (right circuit); $C$ denotes any `reverse
communication' degrees-of-freedom which enter the black hole interior.
[Note, the initial joint quantum state of $(B,N,R)$ is arbitrary.]
  \label{Hawking_process}}
\end{figure}

Based on the no-communication decomposition theorem \cite{Werner} (see
Fig.~\ref{Hawking_process}a) any generic unitary process acting across a
horizon and generating black hole radiation can be decomposed into a
pair of unitary subprocesses which act exclusively on either side of the
horizon \cite{tensor} (see Fig.~\ref{Hawking_process}b).  We label those
{\it external\/} degrees-of-freedom in the black hole's neighborhood as
subsystem $N$ before the action of the externally-acting unitary, $W$,
and those that remain in this neighborhood afterwards as subsystem $N'$.
Similarly, the operation $V$, acting exclusively on the internal
degrees-of-freedom, maps subsystems $B$ and $C$ to subsystem $B'$. Here
subsystem $C$ is included to represent the possibility of a `reverse
communication' channel as allowed from the no-communication
decomposition theorem (see Fig.~\ref{Hawking_process}a).

We may use the circuit shown in Fig.~\ref{Hawking_process}b to represent
the generation of any amount of black hole radiation, which we label
subsystem $R'$. Thus, we might consider using $R'$ for the production of
each individual radiated quanta. Similarly, we might instead consider
using it for the entire radiation generated during some long epoch of a
black hole's lifetime. We shall take this latter approach.

Further, we note that Fig.~\ref{Hawking_process}b is not to be
interpreted as a spacetime diagram. In particular, we do not require
that there is any space-like hypersurface which simultaneously cuts
through the subsystems there displayed. For example, we do not require
that subsystem $C$ all arrives in one block for processing inside the
black hole by the unitary operation $V$. 
From this perspective, a quantum
circuit diagram is a very flexible and powerful construct (see the
Supplementary Information for a generalization that includes arbitrary
infallen matter).


Given Fig.\ref{Hawking_process}b, we can provide equivalent
descriptions of the global Hilbert space,
\begin{eqnarray}
{\cal H}_{\text{global}}
&\cong&{\cal H}_B \otimes {\cal H}_N \otimes {\cal H}_R \otimes
{\cal H}_{\text{extra}} \nonumber \\
&\cong &{\cal H}_B \otimes {\cal H}_C \otimes 
{\cal H}_{N'}\otimes {\cal H}_{R'} \otimes {\cal H}_{R}
\otimes {\cal H}_{\text{extra}} \nonumber \\
&\cong &{\cal H}_{B'} \otimes {\cal H}_{N'}
\otimes {\cal H}_{R'} \otimes {\cal H}_{R}
\otimes {\cal H}_{\text{extra}}, 
\end {eqnarray}
where ${\cal H}_{\text{extra}}$ are any extra degrees-of-freedom not
otherwise mentioned. These different decompositions simply reflect the
fact that subsystem $B'$ is unitarily related to the joint subsystem
$(B,C)$.  Similarly, subsystem $N$ is unitarily related to the joint
subsystem $(C,N',R')$.  These relationships and
Fig.~\ref{Hawking_process}b then lead to the following intuitive
observation: Any entanglement that exists between late and early epoch
radiation, $R'$ and $R$, must have come from entanglement between $N$
and $R$, since entanglement cannot be created non-locally. We shall see
that this intuition places powerful constraints on the compatibility of
our three axioms.

In order to obtain a contradiction from this intuition, we start by
introducing the quantum mutual information
\begin{equation}
S(X:Y)\equiv S(X)+S(Y)-S(X,Y),
\end{equation}
which provides a measure of correlations (both quantum and classical)
between subsystems $X$ and $Y$.  Here $S(X)$ is the von Neumann entropy.
Now strong subadditivity
\begin{equation}
S(X,Y,Z)+S(X)\le S(X,Y)+S(X,Z)
\label{SS}
\end{equation}
of the von Neumann entropy has been used before in the context of black
holes \cite{Mathur09,AMPS}. Adding $S(Z)$ to both sides and rearranging
the terms allows us to rewrite it more conveniently in terms of the
quantum mutual information \cite{Wilde}
\begin{equation}
S(X:Z)\le S(X,Y:Z).
\label{decQMI}
\end{equation}
(By comparison the analogous relation for von Neumann entropies,
$S(X) \le S(X,Y)$, is untrue.)

It is now easy to see that
\begin{equation}
S(R':R)\le S(N:R).
\label{MainResult}
\end{equation}
This only requires observing that subsystems $(C,N',R')$ and $N$ are
unitarily related, as noted above so that $S(C,N',R':R)\equiv S(N:R)$,
then Eq.~(\ref{MainResult}) follows as a straight-forward application
of Eq.~(\ref{decQMI}).  Note that the inequality~(\ref{MainResult}) only
involves correlations between external degrees-of-freedom and hence
relates quantities which are, in principle, directly observable. This
inequality forms our main analytic result and is just one consequence of
Fig.~\ref{Hawking_process} which describes an arbitrary
classically-causal and unitary radiation production process.  In order
to understand its implications we now consider what black hole physics
tells us about the quantities involved.

{\it Implications from the weak correspondence principle:}---In studying an
evaporating black hole one typically relies on Hawking's insight that an
observer might be expected to see a low energy (possibly even the
vacuum) state as she freely falls past the horizon. Unfortunately, she
cannot report any behavior she finds once past the horizon. It therefore
makes sense to construct a criterion for quantum black holes that is
based solely on the {\it external\/} degrees-of-freedom (here labeled
$N$ and $N'$ for the respective states prior and posterior to the
evaporation epoch of interest). Indeed, these external subsystems might
be expected to display entanglement across the horizon (as they would
across any boundary \cite{Braunstein13}).  Therefore we cannot
necessarily assume their having low entropy.

Instead we will argue that the external state seen by the
infalling observer must be at best weakly correlated with the radiation
subsystem. Indeed, this holds identically for any pure state like the
vacuum. Further, unlike the entropy, this property is inherited by any
subsystem such as the external degrees-of-freedom.  We therefore assume
that there exists a parameter $\eta\ll 1$, such that
\begin{equation}
S(N:R) 
\le \eta\, S_{\text{BH}},
\label{EPeqn}
\end{equation}
where $S_{\rm BH}$ is the black hole's {\it initial\/} thermodynamic
entropy and we henceforth set the Boltzmann constant to one.  Should
this condition fail, an infalling observer would see an incredibly mixed
state (e.g., a near uniform mixture of roughly $10^{10^{77}}$ orthogonal
quantum states for an initially stellar mass black hole) as she
approached the horizon and correspondingly huge energies.

Indeed, we can easily estimate the parameter $\eta$ based on 't Hooft's
entropic bound \cite{Hooft}. He showed that if one excludes
configurations of ordinary matter whose energies are so large that they
inevitably undergo gravitational collapse, one finds
$S_{\text{matter}}\le A^{3/4}$ (where $A$ is the surface area
surrounding the matter in Planck units).  Therefore, excluding the
possibility of exotic matter with super-ordinary entropic content in the
external neighborhood implies that $\eta \simeq
(\mu^3/S_{\text{BH}})^{1/4}$, where the surface area of the external
neighborhood is chosen to be $\mu$ times that of the original black
hole. The condition that $\eta\ll 1$ is easily satisfied for any but the
most microscopic initial black holes (even for large neighborhoods,
e.g., $\mu=10^4$). Thus, were our condition~(\ref{EPeqn}) to fail, a
quantum black hole's exterior would be radically different from its
theoretical classical counterpart. (See the Supplementary Information
for a generalization of the weak correspondence principle.)

{\it Implications from unitarily generated radiation:}---Returning now
to our analysis of the terms in Eq.~(\ref{MainResult}) let us consider
the correlations between the late and early epoch radiation
\cite{Page93,Mathur09,AMPS}. As already noted in the introduction, the
general consensus today is that for unitarity to be preserved a black
hole must evaporate completely as radiation. As such, the quantum state
of the sum total of the radiation should be pure and hence have zero net
von Neumann entropy.  Our best understanding of the growth and decay of
the (von Neumann) entropy of black hole radiation comes from models
based on random unitaries \cite{Page93,Braunstein13}.  Within such
models, the entropy of radiation grows monotonically (at almost exactly
the maximal rate of one bit's worth of entropy per qubit of radiation)
until the Page time. From the Page time onwards the entropy in the net
radiation (i.e., including the radiation from the early epoch)
monotonically decreases at the same rate, reaching a net value of zero
when the black hole has evaporated away.  Because the global state of
the net radiation is pure, the early epoch radiation and late epoch
radiation are (nearly maximally) entangled with each other; each
carrying almost exactly one-half the entropy of the {\it initial\/}
black hole ($S_{\text{BH}}$). The huge dimensionalities involved
guarantee that this monotonic rise and fall of the radiation's entropy
must constitute the generic behavior for any unitarily evaporating black
hole \cite{Braunstein13}.

{\it Paradox:}---Consider now the scenario where we follow a black hole
to a relatively late stage of its complete evaporation.  In particular,
when its area has shrunk to some small fraction of its original size,
but is still much larger than the Planck scale so the decomposition shown
in Fig.~\ref{Hawking_process}b should still hold true and Planck scale
physics may be excluded. In this case the net radiation so far can be
split into pre-Page time radiation $R$ and post-Page time radiation $R'$
(produced up until the black hole has reached this specified fraction of
its original area). Consequently, the joint state $(R',R)$ will be very
nearly pure so that $S(R',R)$ is negligible, but $S(R') \simeq
S(R)\simeq \frac{1}{2} S_{\text{BH}}$. More precisely, we suppose there
exists a parameter $\varepsilon \ll 1$ such that
\begin{equation}
S(R':R)\ge (1-\varepsilon)\, S_{\text{BH}}.
\label{RadENT}
\end{equation}

We now easily see that we have a paradoxical situation by combining
Eq.~(\ref{RadENT}) with Eq.~(\ref{EPeqn}) into
Eq.~(\ref{MainResult}) yielding
\begin{equation}
1\le \varepsilon+\eta\ll 1,
\end{equation}
whatever the details of the radiation process. (The relationship between
this result and the firewall paradox is explored in the Supplementary
Information as is its generalization to include infallen matter.)

{\it Discussion:}---We must accept one (or more) of: (a) black holes
cannot unitarily evaporate away completely; (b) there is a violent
failure of the correspondence principle;
(c) there is at least a weak violation of the classical causal structure
of a quantum black hole since a black hole's horizon can allow (quantum)
information out.

Both options (a) and (b) have observational consequences.  Indeed, any
loss of unitarity would infect almost every other quantum mechanical
process \cite{Hooft}.  Similarly, a failure of the correspondence
principle would imply either the existence of exotic matter surrounding
a black hole, or its atmosphere would need to extend to truly enormous
proportions implying that black holes would in fact not be compact
objects.

By comparison, option (c) only requires that the Hilbert space
within the black hole `leak away' --- we would call such a mechanism
tunneling \cite{Braunstein11,Parikh00,Braunstein13} --- and we would
conclude that the black hole radiation originates from {\it within\/}
the horizon \cite{baryon}.  Option (c) therefore appears to be the least
extreme choice among the possible resolutions, having no direct
observational consequences other than allowing for the preservation of
unitarity and the correspondence principle.

Our version of the paradox precludes any difficulties with information
preservation due to the possible existence of a singularity within the
black hole by: i) presuming unitarity from the outset; and ii) only
investigating the physics of black holes external to the horizon.  This
is both the strength and the weakness of our approach as it implies that
our putative resolution literally scratches the surface of the original
information paradox.

The authors thank D.\ Harlow, P.\ Kok, S.\ Massar, S.\ Mathur and J.\
Oppenheim, for fruitful discussions.

\subsection*{SUPPLEMENTARY INFORMATION}

\subsection*{The firewall argument recast}

Let us express the key entropic strong subadditivity inequality,
Eq.~(\ref{SS}), from the firewall paper \cite{AMPS} in terms of our
notation from Fig.~\ref{Hawking_process}b. We immediately obtain
\begin{equation}
S(R,R') +S(C,R')\ge S(R')+S(C,R',R), 
\end{equation}
or equivalently as
\begin{equation}
S(R':R)\le S(C,R':R). 
\label{firewall}
\end{equation}
The argument then goes that with regard to the specific mechanism of
pair creation into vacuum, the joint subsystem $(C,R')$ should be pure,
signaling the so-called ``no drama'' scenario of an infalling observer,
and hence the right-hand-side of Eq.~(\ref{firewall}) should be zero.
However, the left-hand-side of this equation can be as large as
the black hole's initial entropy, $S_{\rm BH}$, implying a contradiction.

Crucially, this argument relies explicitly on i) pair creation into
vacuum as the mechanism of radiation; and hence ii) ignoring any
explicit role of the external neighboring degrees-of-freedom, $N$ and
$N'$, from participating in the long-term dynamics of the evaporating
black hole.  As we noted in our introduction, wanting to escape from the
former assumption was the motivation for the present paper. It is the
latter assumption, however, which reveals a critical flaw in the
firewall argument.

In particular, we now know that entanglement across any boundary (such
as the horizon) is the natural consequence of the quantum fields across
that boundary being in any non-singular (finite-energy) configuration
\cite{Braunstein13}. Further, it is now widely accepted that the
trans-horizon entanglement of any quantum fields whose quantum state is
not too far from the vacuum will scale as the area of the horizon 
\cite{Eisert10}. Thus, as a black hole evaporates, part-and-parcel of
a complete description of the overall dynamics {\it must\/} include
the dynamics that proportionally decreases the trans-horizon entanglement.
The Hawking process fails to do this. Therefore, we cannot have confidence
that it alone describes the long-term dynamics of an evaporating
black hole. 

In particular, although for a static horizon, such as the Rindler
horizon, these external neighborhood modes completely decouple from the
Hawking radiation, this is simply not possible for a non-static,
shrinking horizon, such as in the scenario of an evaporating black hole.
In this case, we can no longer rely on the intuition from the Hawking
process that $N$ and $N'$ will remain unentangled from the joint
subsystem $(C,R')$. Thus, the existence of a firewall at the horizon
would appear to require extra assumptions such as an initially finite
Hilbert space dimensionality for the black hole interior which
consequently shrinks during the evaporation process \cite{Braunstein13}.

Despite this difficulty, let us recast the firewall argument in a manner
that overcomes both of the limiting assumptions mentioned above. Relying
on the left-hand circuit of Fig.~\ref{Hawking_process}b only we easily
see that
\begin{equation}
S(R':R)\le S(B,N:R),
\label{newfirewall}
\end{equation}
where we use Eq.~(\ref{decQMI}) and the fact that the joint subsystems
$(B',N',R')$ and $(B,N)$ are unitarily related, so that
$S(B',N',R':R)\equiv S(B,N:R)$ (see Fig.~\ref{Hawking_process}b).  Now
taking the conditions leading to Eq.~(\ref{RadENT}), we find that the
near maximal entanglement between post- and pre-Page time radiation,
$R'$ and $R$, respectively, must have come from near maximal
entanglement between early epoch radiation, $R$, and the joint subsystem
$(B,N)$ describing the degrees-of-freedom {\it within and surrounding\/}
the black hole at the Page time. Note that unlike
Eq.~(\ref{MainResult}), Eq.~(\ref{newfirewall}) does not require that
the horizon be classically causal as it does {\it not\/} impose the
no-communication decomposition.

Combining Eqs.~(\ref{RadENT}) and~(\ref{newfirewall}) would then imply
that the joint subsystem $(B,N)$ (i.e., when tracing out early epoch
radiation, $R$) must be in an incredibly mixed state (e.g., a near
uniform mixture of roughly $10^{10^{77}}$ orthogonal quantum states for
an initially stellar mass black hole). Thus, an infalling observer will
necessarily encounter significant ``drama''. However, the problem is
that this analysis cannot specify the location of this drama; it may
occur anywhere in the joint $(B,N)$ subsystem. The most natural
assumption would be that it occurs near the metric singularity well
within the black hole interior; where an infalling observer into even a
theoretical classical black hole fully expects to encounter drama; and
presumably where the infalling partners to the Hawking radiation have
ended up.  Thus, when we explicitly incorporate the external
degrees-of-freedom surrounding the black hole, $N$ and $N'$, the force
behind the generic firewall argument is dissipated.

\subsection*{Generalizing weak correspondence}

We may extend the weak correspondence principle to read, there exists a
parameter $\eta\ll 1$ such that
\begin{equation}
S(N:{\text {Ref}})\le \eta\, S_{\rm BH},
\end{equation}
where `Ref' denotes any distant reference subsystem and $S_{\rm BH}$ is
the black hole's initial thermodynamic entropy. The reasoning is
identical to that used to support Eq.~(\ref{EPeqn}). As an example of
the utility of this generalization let us use it to examine any theory
which claims that information about infallen matter may be stored
externally of the event horizon
\cite{Hooft,Susskind93,Raine94,Susskind95,Mathur09b}.

To do so, let us first recall that for an initially collapsed large
black hole, the information content of the matter that formed it can be
as high as $O(S_{\rm BH}^{3/4})$ from 't Hooft's entropic bound
\cite{Hooft}. Let us further consider a scenario where as the black hole
slowly evaporates we replace each Hawking quanta with another quanta of
infallen information. We may keep this process up for a duration as long
as the Page time without the information about the infallen matter being
radiated away \cite{Braunstein13,Page93}.  By this procedure, we have
created a black hole with the same thermodynamic entropy, $S_{\rm BH}$,
as the original black hole, yet where the infallen matter itself has
entropy $O(S_{\rm BH})$.

Without loss of generality, we may suppose that the infallen matter's
entropy comes from it being entangled with some distant external
reference subsystem `Ref'. In this case, we may immediately invoke our
generalized weak statement of the correspondence principle to argue that
the black hole so constructed, cannot look anything like a classical
black hole in any theory where information about the infallen matter is
stored (or accessible) externally. Thus, for example, black hole
complementarity proposes to have two copies of the information about
infallen matter; one that is accessible only internally and a second
copy externally \cite{Susskind93,Susskind95}. Similarly, fuzzball
complementarity assumes that the information about infalling matter
resides in a fuzzball structure that lies outside the horizon
\cite{Mathur09b}.  However, for black holes as constructed above, the
information content is so large that the external region carrying this
information can only consist of a super-entropic exotic matter or more
likely the horizon must expand to encompass this information.  Thus,
{\it any\/} theory which claims to store information about infallen
matter externally \cite{Hooft,Susskind93,Raine94,Susskind95,Mathur09b}
will necessarily imply an extreme failure of the correspondence between
quantum and classical black holes; presumably violating its whole {\it
raison d'\^{e}tre}.

\subsection*{Generics of black hole radiation}

In the manuscript we considered a black hole with thermodynamic entropy
$S_{\text{BH}}$ which can completely evaporate into a net pure state of
radiation. As discussed, the generic evaporative dynamics of such a
black hole may be captured by the random sampling of subsystems from an
initially pure state consisting of $S_{\text{BH}}$ qubits
\cite{Page93,Braunstein13}. This either neglects infallen matter or
assumes it is pure.

In order to extend our analysis to include infallen matter carrying some
(von Neumann) entropy $S_{\text{matter}}$, we need only take the
initially pure state used above and replace it with a bipartite pure
state consisting of two subsystems: $S_{\text{BH}}$ qubits to represent
the degrees-of-freedom that evaporate away as radiation; and a reference
subsystem. Without loss of generality, the matter's entropy may be
treated as entanglement between these two subsystems, however, here we
shall simplify our analysis by assuming uniform entanglement between the
black hole subsystem and $S_{\text{matter}}$ reference qubits. The
generic properties of the radiation may then again be studied by random
sampling the former subsystem to simulate the ejection of radiation
\cite{Braunstein13}.

The behavior is generic and for our purposes may be summarized in terms
of the radiation's von Neumann entropy, $S(R)$, as a function of the
number of qubits in this radiation subsystem. One finds
\cite{Braunstein13} that $S(R)$ initially increases by one qubit for
every extra qubit in $R$, until it contains
$\frac{1}{2}(S_{\text{BH}}+S_{\text{matter}})$ qubits. From that stage
on it decreases by one qubit for every extra qubit in $R$ until it drops
to $S_{\text{matter}}$ when $R$ contains $S_{\text{BH}}$ qubits and the
black hole has completely evaporated.

Because the von Neumann entropy of a randomly selected subsystem only
depends on the size of that subsystem, the same behavior is found whether
$R$ above represents the early or late epoch radiation with respect
to any arbitrary split. Further, in the simplest case where we choose
the joint radiation $(R,R')$ to correspond to the net radiation from
a completely evaporated black hole we may immediately write down the
generic behavior for the quantum mutual information $S(R':R)$.

In particular, $S(R':R)$ starts from zero when $R$ consists of zero
qubits.  From then on, it increases by two qubits for every extra qubit
in $R$ until $S(R':R)$ reaches $S_{\text{BH}}-S_{\text{matter}}$ when
$R$ contains $\frac{1}{2}(S_{\text{BH}}-S_{\text{matter}})$ qubits. From
that stage on until $R$ contains
$\frac{1}{2}(S_{\text{BH}}+S_{\text{matter}})$ qubits $S(R':R)$ remains
constant, after which $S(R':R)$ decreases by two qubits for every extra
qubit in $R$ until it drops to zero once the $R$ contains the full
$S_{\text{BH}}$ qubits of the completely evaporated black hole
\cite{Braunstein13}. Interestingly, it is during region where $S(R':R)$
is constant that the information about the infallen matter becomes
encoded into $R$ for the first time \cite{Braunstein13}. Finally, setting
$S_{\text{matter}}$ to zero gives the `standard' behavior for $S(R)$ and
$S(R':R)$ upon which the results in the manuscript are derived.

\subsection*{Including infallen matter}

\begin{figure}[h!]
\includegraphics[width=50mm]{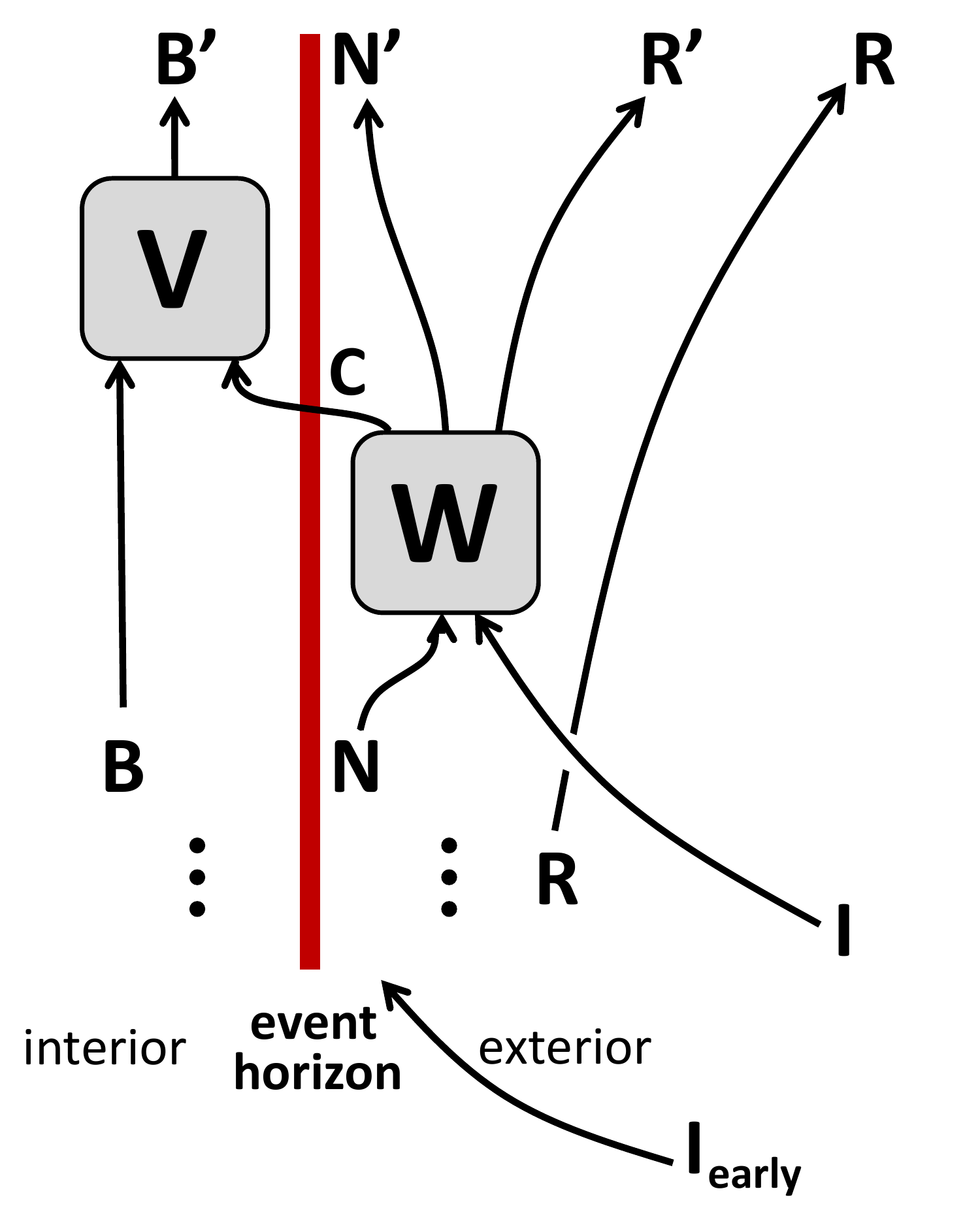}
\vskip -0.1in
  \caption{
Quantum circuit diagram for evaporation of a quantum black hole with
causal horizon and infallen matter.  Subsystem $I$ denotes infallen
matter that falls into the region surrounding the black hole to
participate in late epoch radiation generation.  (This does not exclude
the possibility that the matter falls directly into the black hole.)
Subsystem $I_{\text{early}}$ denotes matter infalling at earlier times
or even that collapses to form the original black hole.
\label{Fig2}
}
\end{figure}

In the main body of the manuscript we did not explicitly include entropy
associated with infallen matter.  Fig.~\ref{Fig2} shows the most general
scenario. Subsystem $I$ denotes the matter that falls into the region
surrounding the black hole where radiation is produced. Thus, we suppose
that late epoch radiation can in principle come from the joint subsystem
$(N,I)$. In this figure we also include subsystem $I_{\text{early}}$
which denotes matter that has fallen into the region surrounding the
black hole at an earlier epoch or indeed matter that may have collapsed
to form the black hole in the first place.

We start as before applying strong subadditivity expressed in terms
of quantum mutual information
\begin{equation}
S(R':R) \le S(C,N',R':R)
= S(N,I:R) = S(N:R),
\label{NewMain}
\end{equation}
Here, we use the fact that joint subsystems $(C,N',R')$ and $(N,I)$ are
unitarily related. Finally, the most natural assumption is that the
infallen matter $I$ is independent of the quantum state of the black
hole, $(B,N)$, or its early epoch radiation $R$. The originally
derived inequality of Eq.~(\ref{MainResult}) is therefore found to still
hold in the presence of infallen matter.

From the analysis including infallen matter given in the
previous section we have enough to complete our analysis.  As in the
manuscript, we take $R$ to be all the early epoch radiation until the
Page time, and we let $R'$ denote all the radiation generated from the
Page time onwards until the black hole has shrunk to a size much smaller
than the original black hole, but still much larger than the Planck
scale. In this case, instead of Eq.~(\ref{RadENT}), we suppose that
there exists a parameter $\varepsilon\ll 1$ such that
\begin{equation}
S(R':R)\ge (1-\varepsilon) \, S_{\text{BH}}-S_{\text{matter}},
\end{equation}
where $S_{\text{matter}}\equiv S(I_{\text{early}},I)$ is the net entropy
contained in all the infallen matter.  Next, combining this with our
weak correspondence principle, Eq.~(\ref{EPeqn}), we find
\begin{equation}
1 -\frac{S_{\text{matter}}}{S_{\text{BH}}} \le \epsilon + \eta \ll 1.
\end{equation}
Once again we obtain a contradiction except in the extreme and
unphysical case of a black hole whose infallen matter contains virtually
as much entropy as the entire black hole's Bekenstein-Hawking entropy
$S_{\text{BH}}$. Thus, the conclusions given in the manuscript hold
even when the analysis is extended to include infallen matter.


\begin{thebibliography}{99}

\bibitem{Hawking76}
S.\ W.\ Hawking,
Phys.\ Rev.\ D {\bf 14}, 2460 (1976).

\bibitem{Braunstein13} 
S.\ L.\ Braunstein, arXiv:0907.1190v1 (2009); published as
S.\ L.\ Braunstein, S.\ Pirandola and K.\ \.Zyczkowski,
Phys.\ Rev.\ Lett.\ {\bf 110}, 101301 (2013);

\bibitem{Mathur09} S.\ D.\ Mathur,
Classical Quantum Gravity {\bf 26}, 224001 (2009).

\bibitem{AMPS}
A.\ Almheiri, et al., 
J.\ High Energy Phys.\ {\bf 02} (2013) 062.

\bibitem{Braunstein07}
S.\ L.\ Braunstein and A.\ K.\ Pati,
Phys.\ Rev.\ Lett.\ {\bf 98}, 080502 (2007).

\bibitem{Hooft}
G.\ 't Hooft, in {\it Salamfestschrift: A Collection of Talks},
edited by A.\ Ali, J.\ Ellis, and S.\ Randjbar-Daemi (World
Scientific, Singapore, 1993), Vol.\ 4, p.~284.

\bibitem{Bekenstein94}
J.\ D.\ Bekenstein,
Phys.\ Rev.\ D {\bf 49}, 1912 (1994).

\bibitem{Giddings95}
S.\ B.\ Giddings,
Phys.\ Rev.\ D {\bf 51}, 6860 (1995).

\bibitem{Banks84}
T.\ Banks, et al.,
Nucl.\ Phys.\ {\bf B244}, 125 (1984).

\bibitem{Werner}
T.\ Eggeling, D.\ Schilingemann and R.\ F.\ Werner,
Europhys.\ Lett.\ {\bf 57}, 782 (2002).

\bibitem{tensor}
The tensor-product structure between interior and exterior is
guaranteed by existence of the horizon, see Ref.~\onlinecite{Braunstein13}.

\bibitem{Wilde}
M.\ M.\ Wilde,
{\it Quantum Information Theory\/}
(Cambridge University Press, Cambridge, 2013) p.~270.

\bibitem{Page93}
D.\ N.\ Page,
Phys.\ Rev.\ Lett.\ {\bf 71}, 3743 (1993).

\bibitem{Parikh00}
M.\ K.\ Parikh and F.\ Wilczek,
Phys.\ Rev.\ Lett.\ {\bf 85}, 5042 (2000).

\bibitem{Braunstein11}
S.\ L.\ Braunstein and M.\ K.\ Patra,
Phys.\ Rev.\ Lett.\ {\bf 107}, 071302 (2011).

\bibitem{baryon}
This would also naturally solve the difficulty of extreme baryon
non-conservation in black hole evaporation; see Ya.\ B.\ Zeldovich,
Sov.\ Phys.\ JETP {\bf 45}, 9 (1977).


\bibitem{Eisert10}
J.\ Eisert, et al., 
Rev.\ Mod.\ Phys.\ {\bf 82}, 277 (2010).

\bibitem{Susskind93}
L.\ Susskind, L.\ Thorlacius and J.\ Uglum,
Phys.\ Rev.\ D {\bf 48}, 3743 (1993).

\bibitem{Raine94}
D.\ J.\ Raine and D.\ W.\ Sciama,
J.\ Stat.\ Phys.\ {\bf 77}, 223 (1994).

\bibitem{Susskind95}
L.\ Susskind,
J.\ Math.\ Phys\. {\bf 36}, 6377 (1995).

\bibitem{Mathur09b}
S.\ D.\ Mathur,
Int.\ J.\ Mod.\ Phys.\ {\bf 18}, 2215 (2009).

\end{thebibliography}
\end{document}